# A numerical study of the run-up and the force exerted on a vertical wall by a solitary wave propagating over two tandem trenches and impinging on the wall

by

G.A. Athanassoulis ([1]) ([a]), C.P. Mavroeidis ([1]) ([b]), P.E. Koutsogiannakis ([1]) ([c]), Ch.E. Papoutsellis ([2]) ([d])

([1]) School of Naval Architecture and Marine Engineering, National Technical University of Athens (NTUA), Zografos, Athens, Greece

([2]) Aix Marseille Univ, CNRS, IRD, INRA, Coll France, CEREGE, Aix-en-Provence, France

([a]) mathan@central.ntua.gr  ([b]) con.mavroeidis@gmail.com
([c]) pkoutsogian@hotmail.com  ([d]) cpapoutsellis@gmail.com

## Table of contents

# A numerical study of the run-up and the force exerted on a vertical wall by a solitary wave propagating over two tandem trenches and impinging on the wall

**Abstract**

The propagation and transformation of water waves over varying bathymetries is a subject of fundamental interest to ocean, coastal and harbor engineers. The specific bathymetry considered in this paper consists of one or two, naturally formed or man-made, trenches. The problem we focus on is the transformation of an incoming solitary wave by the trench(es), and the impact of the resulting wave system on a vertical wall located after the trench(es). The maximum run-up and the maximum force exerted on the wall are calculated for various lengths and heights of the trench(es), and are compared with the corresponding quantities in the absence of them. The calculations have been performed by using the fully nonlinear water-wave equations, in the form of the Hamiltonian coupled-mode theory, recently developed in Papoutsellis et al (Eur. J. Mech. B/Fluids, Vol. 72, 2018, pp. 199–224). Comparisons of the calculated free-surface elevation with existing experimental results indicate that the effect of the vortical flow, inevitably developed within and near the trench(es) but not captured by any potential theory, is not important concerning the frontal wave flow regime. This suggests that the predictions of the run-up and the force on the wall by nonlinear potential theory are expected to be nearly realistic. The main conclusion of our investigation is that the presence of two tandem trenches in front of the wall may reduce the run-up from (about) 20% to 45% and the force from 15% to 38%., depending on the trench dimensions and the wave amplitude. The percentage reduction is greater for higher waves. The presence of only one trench leads to reductions 1.4 – 1.7 times smaller.

**Keywords:** nonlinear water waves; wave-trench interaction; solitary wave over varying bathymetry; run-up on vertical wall; force on vertical wall; submerged breakwater

# 1. Introduction

The propagation and transformation of water waves over varying bathymetries is a subject of fundamental interest to ocean, coastal and harbor engineers. A specific bathymetry that has attracted much interest in the last decades is that consisting of one or more, naturally formed or man-made, trenches, also called cavities or pits. As an incident wave propagates over such a bathymetry, the interaction of waves with the seabed may seriously deform the free surface and cause various hydrodynamic effects as, for example, wave reflection and scattering, having potential impact on natural or engineered systems. Man-made trenches are often shaped near or in front of harbors by dredging works, in order to provide fill rubble for breakwater caissons or navigation channels for big ships (H. S. Lee and Kim 2004; H. S. Lee et al. 2009). It has also been recognized that a single trench or a system of multiple trenches modify the local wave climate and may function as a submerged breakwater, called pit breakwater or multiple-pit breakwater, respectively; see e.g. (McDougal et al. 1996; H. S. Lee 2004; H. S. Lee and Kim 2004; H. S. Lee et al. 2009; Kim et al. 2015). Another issue related with such types of bathymetries is that the trenches (cavities) usually act as sediment pockets or traps, resulting in a strong density stratification and the development of internal waves within them (Lassiter 1972; Ting 1994). Although the latter issue is out of the scope of the present work, we mention here that the motion of sediment particles is strongly affected by the vortices which are developed at the edges and within the trench (Chang and Lin 2015; Wu et al. 2015a, b).

The vast majority of early works studying the wave-trench interaction have been conducted in the context of linear wave theory, either shallow water or full-linear potential theory([1]). Probably the earliest systematic study in this context is due to Lassiter (1972), who formulated the two-dimensional (2D) scattering problem in terms of complementary variational integrals of Schwinger's type, and presented numerical calculations for the complex reflection and transmission coefficients of a normally incident monochromatic wave, passing over a rectangular trench containing two different fluids and having different depths in the upstream and the downstream sides. The interaction of normally or obliquely incident water waves with a rectangular submarine trench has been investigated by several authors in the 80s and 90s; see e.g. J.-J. Lee and Ayer (1981), Miles (1982), Kirby and Dalrymple (1983), Ting and Raichlen (1986), Kirby et al. (1987) and Williams (1990), usually restricted to 2D geometries. A survey of some methods, for solving the problem of linear wave transformation by 2D bathymetric anomalies, has been conducted by Bender and Dean (2003), who studied the reflection and transmission of normally incident waves by trenches and shoals with sloped transitions. The analysis was later extended to 3D geometries, using the full-linear theory, by Williams and Vazquez (1991). McDougal et al. (1996) extended the shallow-water wave analysis of Williams (1990) to the 3D case with multiple pits. Some authors studied the same problem by using the mild-slope or the modified mild-slope equation, providing analytical or semi-analytical solutions to reflection and transmission coefficients for typical trenches, valid in the linear, long-wave regime; see e.g. (Jung et al. 2008; Michalsen et al. 2008; Xie et al. 2011; Xie and Liu 2012; H.-W. Liu et al. 2013). Recent investigations, in the context of the full-linear potential theory, have been presented by Chakraborty and Mandal (2014, 2015), for normal and oblique incidence of waves over a single rectangular trench, respectively, and by Kar et al. (2018), where a system of two 2D trenches of general shape has been considered. See also Roy et al. (2017) and Chapter 3 of the recent book by Mandal and De (2015).

Despite the extensive investigation of the wave transformation over (one or more) trenches conducted by using linear wave theories, analogous works using nonlinear theories are much more limited. Our main concern in this paper is the study of the transformation of a solitary wave propagating through a homogeneous liquid and passing over two tandem trenches, and the impact of the resulting wave system on a vertical wall. Incidentally, we also study the same problem with a single trench. To the best of our knowledge, only the latter case has been considered in the existing literature; see (Chang et al. 2011, 2012; Chu et al. 2015; Chang 2019). The emphasis, in the above-mentioned works, lies in the solitary wave transformation by the trench and on the study of the vortical fluid motion induced by the trench. For this purpose, the above authors utilized a 2D viscous flow model, based on stream function and vorticity formulation, implemented by means of the finite analytic numerical method (Chen and Chen 1984). Clearly, there are two different types of complicacies in studying the interaction of a solitary wave with trenches and the impact of the resulting wave system on a vertical wall. First, the flow becomes vortical within and near the trenches (especially at the edges), a feature which cannot be grasped by means of the potential theory, either linear or nonlinear. Second, the impact on the vertical wall is a violent, strongly nonlinear, phenomenon, and the run-up is practically incompatible with the viscous no-slip boundary condition. Fortunately, the detailed, experimental and numerical, analysis of the solitary wave – trench interaction, provided in the above-mentioned papers by Prof. Chang and co-workers, established that the strongly vortical flow regime is restricted within and near the trench and, thus, its effects on the wave flow, especially on the main frontal part, is not important. Accordingly, it seems reasonable to study the problem by using nonlinear potential theory, pointing to the calculation of the surface wave transformation, as well as the run-up and the force exerted on the wall lying behind the trench(es). This is the main purpose of the present paper. Since the depth in the trenches may be two or more times the initial depth, it is not safe to use shallow-water nonlinear theories, which may be invalidated by this deepening. Our approach utilizes a fully nonlinear Hamiltonian coupled-mode theory (HCMT), developed recently by two of the present authors; see (Athanassoulis and Papoutsellis 2017; Papoutsellis and Athanassoulis 2017; Papoutsellis et al. 2018). This theory and its numerical implementation are valid for any depth, any (smooth) bathymetry, however steep, and any level of nonlinearity (steepness) of non-breaking surface waves. The apparent incompatibility between the rectangular trench bathymetry and the smoothness requirements of HCMT, is resolved by approximating the rectangular trench(es) as smooth, nearly-rectangular cavities, with the same length (width) and height (depth) as the rectangular ones and very steep lateral boundaries, using a combination of tanh-functions. This change has negligible effects on the free-surface elevation and the main wave flow, although may lead to serious locally, near the trenches.

---
([1]) That is, linear potential theory valid for arbitrary water depth.

In the context of the fully nonlinear potential theory, the primitive unknown fields are the free-surface elevation $\eta(x,t)$ and the wave potential $\Phi(x,z,t)$. The latter is defined in the unknown ($\eta$ – dependent) domain

$$D_h^\eta = D_h^\eta(X,t) = \{(x,z) \in X \times \mathbb{R}, -h(x) < z < \eta(x,t)\}, \quad t \in [t_0, t_1], \tag{1}$$

where $X = [a,b]$ is the common projection of the free surface $\Gamma^\eta: z = \eta(x,t)$ and the seabed topography $\Gamma_h: z = -h(x)$ on the horizontal axis. The right-end vertical boundary, at $x = b$, is the wall on which the propagating wave is reflected, while the left-end vertical boundary, at $x = a$, is located far enough in order to have no influence on the main flow phenomena studied herein. The differential equations, and the boundary and initial conditions applied to the unknown fields $\eta(x,t)$ and $\Phi(x,z,t)$ are given in Sec. 2. The HCMT starts with the following, *exact and rapidly convergent*, series expansion representation of the wave potential:

$$\Phi(x,z,t) = \varphi_{-2}(x,t) Z_{-2}(z;\eta,h) + \varphi_{-1}(x,t) Z_{-1}(z;\eta,h) + \\ + \sum_{n=0}^{\infty} \varphi_n(x,t) Z_n(z;\eta,h), \tag{2}$$

where $\varphi_n = \varphi_n(x,t)$ are unknown modal amplitudes, and $Z_n = Z_n(z;\eta,h)$, $n \geq -2$, are vertical basis functions, explicitly given in terms of the local values of the free-surface elevation $\eta(x,t)$ and bathymetry $h(x)$; exact definitions (explicit formulae) of all functions $Z_n$, $n \geq -2$, are given in Sec. 2.2 of Papoutsellis et al. (2018). The somewhat curious indexing of the first two terms of series (2), $n = -2, -1$, comes from the fact that the remaining terms, with indices $n = 0$ and $n \geq 1$, are conceptually related with the propagating and the evanescent modes of the linear theory. Also, if the HCMT is linearized, $Z_0$ and $Z_n$, $n \geq 1$ become identical with the corresponding vertical modes of the standard linear theory. Nevertheless, it should be stressed herein that, in the context of the HCMT, the derivation of the vertical functions $Z_n$, $n \geq -2$ is completely independent from any linearization argument. The first two terms, with indices $n = -2, -1$, are indirect representations of the vertical derivative of the wave potential on the free surface $z = \eta(x,t)$ and on the bottom $z = -h(x)$ and, for this reason, are referred to as the free-surface mode and the bottom mode respectively.

The mathematical theory behind the series expansion (2) is presented in detail in Athanassoulis and Papoutsellis (2017). As it is shown therein, the infinite series expansion (2) is not an approximate or asymptotic representation. Instead, it is an *exact and rapidly convergent semi-separation of variables* in the irregular, instantaneous fluid domain $D_h^\eta(X,t)$, valid under plausible assumptions on the smoothness (differentiability) of the boundary functions $\eta(x,t)$ and $h(x)$. The most important properties of series (2), justifying its use in the HCMT, are the following:

- The $C^2$ norm of $\varphi_n(x,t)$, as a function of $x \in X$, defined by

$$\|\varphi_n\|_{C^2} = \max\{|\varphi_n(x,t)| + |\partial_x \varphi_n(x,t)| + |\partial_x^2 \varphi_n(x,t)|,\ x \in X\},$$

satisfies the estimate $\|\varphi_n\|_{C^2} = O(n^{-4})$,
- Series (2) can be differentiated term-by-term at least twice, with respect to all independent variables,
- The resulting series, after the term-wise differentiation, converge (rapidly as well) to the corresponding derivatives of $\Phi$, throughout the whole fluid domain $D_h^\eta(X,t)$, up to and including the boundaries.

Rigorous proofs of the above statements are given in Athanassoulis and Papoutsellis (2017), Theorems 1 and 2, and corresponding Corollaries.

By modelling the rectangular trenches as smooth, steep (almost rectangular) cavities, and using representation (2), the fully nonlinear problem is recast as a coupled-mode system of horizontal differential equations with respect to the new unknown fields $\eta, \varphi_{-2}, \varphi_{-1}, \varphi_0, \varphi_1, ...$ (see Sec. 2). The obtained new, exact reformulation of the fully nonlinear problem exhibits two important features: **i)** *it is a fixed-domain problem* (all equations to be solved are now defined in the fixed spatial domain $X = [a,b]$ for all $t \geq t_0$), **ii)** *it is a dimensionally reduced problem*, in the sense that the equations to be solved are all defined in the 1D spatial domain $X = [a,b]$, instead of the 2D domain $D_h^\eta(X,t)$. Of course, the obtained horizontal equations (see Eqs. (8) in Sec. 2) form an infinite system, which is truncated to a finite order, during the numerical implementation of the method (see Sec. 3). This truncation is realized by ignoring the terms of the series (2) (and the corresponding equations) with $n > M$, where $M$ is usually taken to be 3, 4 or 5([2]). The ignored higher-order terms are consistently small, independently of

---

([2]) The required number of $M$ is always found by a preliminary investigation of the convergence of the numerical scheme. In almost all studied cases (herein and in other applications), $M = 5$ suffices for numerical convergence. The how much the boundaries differ from the planar ones and how steep they are. Thus, even the finite-dimensional approximations obtained by the HCMT *are not perturbative*, as regards the boundary shape. This has been clearly demonstrated by comparing our numerical solutions to simple, standard, benchmark problems with known analytical results in Athanassoulis and Papoutsellis (2017). Accordingly, our model fully encompasses the nonlinear and the dispersive characteristics of the wave flow, leaving aside only the vortex formation in the trenches, since it is a potential theory. Nevertheless, our model predicts almost identical results, for the free-surface elevation above and after the trench, with the experimental measurements and the numerical results obtained by Chang et al. (2011), using 2D Navier-Stokes equations; see the comparisons in Sec. 4.4. Thus, we can argue that the HCMT provides reasonably accurate calculations for the main wave flow, far from the trenches. Of course, additional experimental verification is highly desirable.

Regarding its ability to predict the run-up and the force exerted on a vertical wall by a solitary wave propagating over a flat seabed, the HCMT has been thoroughly validated against experimental measurements and numerical results by other fully nonlinear solvers in Papoutsellis et al. (2018). The excellent agreement between HCMT predictions and experimental results for both the run-up and the force can be seen in Figs. 4 and 5, respectively, of the afore-mentioned work.

The main objective of the present paper is the investigation of the effect of trenches on the run-up and the force exerted on a vertical wall by a solitary wave, propagating over and transformed by a bathymetry with trench(es). That is, the present paper continues the investigation of the trenches as submerged breakwaters, moving from the linear to the fully nonlinear theory, when the incident wave is a solitary one. The latter is constructed as a solution to the fully nonlinear water-wave equations as well, by means of a highly-accurate iterative solver of Clamond and Dutykh (2013), which is briefly described in Appendix A. Systematic results for the maximum run-up and the maximum force on the wall, when an incident solitary wave of amplitude $a$ propagates over a reference depth $h_0$, with $a/h_0 = 0.2, 0.3$, 0.4, passes over one or two tandem trenches and impinges on the wall, are presented in Sec. 5. The impact of the trenches on the run-up and the force is nonlinearly dependent on their dimensions and the wave amplitude. The greater the dimensions and the amplitude, the more intense the reduction of the run-up and force becomes, reaching the figures of 45% and 38%, respectively, for the largest trenches considered in this work, and $a/h_0 = 0.4$. When only one of the two trenches is present, the reduction of the run-up and the force is 1.4 – 1.7 times smaller.

## 2. An overview of the fully non-linear Hamiltonian Coupled-Mode Theory

In this section, a concise overview of the classical nonlinear water-wave equations and the HCMT is presented, restricted to the two-dimensional (2D) case, in accordance with the applications we are going to discuss.

### 2.1 Classical differential and variational formulation

Consider a Cartesian coordinate system $Oxz$, with the $x-$axis coinciding with the quiescent free surface, and the $z$-axis pointing upwards. The fluid is assumed homogeneous and incompressible, and the flow inviscid and irrotational. The fluid velocity in the (unknown and time-dependent) domain $D_h^\eta(X,t) = D_h^\eta$, see Eq. (1), is described by means of a wave potential $\Phi = \Phi(x,z,t)$ satisfying the following, well-known, equations (Wehausen and Laitone 1960; Mei et al. 2005, Ch.1):

---

only cases for which $M$ may need to be greater (up to 10) are those with highly nonlinear waves, just before the breaking limit.

$$\Delta \Phi = \left(\partial_x^2 + \partial_z^2\right)\Phi = 0, \qquad \text{in } D_h^\eta, \tag{3a}$$

$$\partial_t \eta + \partial_x \eta \, \partial_x \Phi - \partial_z \Phi = 0, \qquad \text{on } \Gamma^\eta, \tag{3b}$$

$$\partial_t \Phi + \frac{1}{2}|\nabla \Phi|^2 + g\eta = 0, \qquad \text{on } \Gamma^\eta, \tag{3c}$$

$$\partial_x h \, \partial_x \Phi + \partial_z \Phi = 0, \qquad \text{on } \Gamma_h, \tag{3d}$$

where $g$ is the acceleration of gravity. Eqs. (3a) and (3b) are the kinematic and the dynamic free-surface boundary conditions, while Eq. (3d) is the bottom boundary condition. These equations should be supplemented by lateral boundary conditions, which, in the present paper, have the form

$$\partial_x \Phi(x=a,t) = 0 = \partial_x \Phi(x=b,t), \tag{4a,b}$$

modelling vertical walls at the end points $x=a$ and $x=b$. The initial state of the flow, i.e. the initial fields $\eta(x,t_0)$ and $\Phi(x,z,t_0)$, should be also given for any specific application. Let it be noted that the specification of nonlinear initial conditions (initial flow) is not an easy problem for water waves. Such a specification usually requires the solution of another nonlinear problem, by means of some simpler or specific method. For the applications we are going to discuss in this paper, the appropriate initial flow corresponds to the flow of a solitary wave located in the flat-bottom region, far enough from the left wall, $x=a$, and beyond the first trench. Its derivation is briefly presented in Appendix A.

The nonlinear problem described by Eqs. (3) and (4) admits of an unconstrained variational formulation, introduced by Luke (1967); see also Whitham (1974), Ch. 13. The action functional of Luke's Variational Principle has the form

$$\mathcal{S}[\eta,\Phi] = \int_{t_0}^{t_1}\int_X L(\eta,\Phi)\,d\boldsymbol{x}\,dt, \tag{5a}$$

where

$$L(\eta,\Phi) = \int_{-h}^{\eta}\left(\partial_t\Phi + \frac{1}{2}\{(\partial_x\Phi)^2 + (\partial_z\Phi)^2\} + gz\right)dz. \tag{5b}$$

Luke's Variational Principle states that the fields $\eta$ and $\Phi$ satisfy Eqs (3) and (4) if and only if they render the action functional $\mathcal{S}[\eta,\Phi]$ stationary, that is, they satisfy the variational equation

$$\delta\mathcal{S}[\eta,\Phi;\delta\Phi,\delta\eta] = \delta_\Phi\mathcal{S}[\eta,\Phi;\delta\Phi] + \delta_\eta\mathcal{S}[\eta,\Phi;\delta\eta] = 0. \tag{6}$$

An important feature of the variational formulation (of any problem) is that it facilitates the reformulation of the problem, by using various representations of the unknown fields, involved in the functional, in terms of other, more convenient ones, and performing the corresponding variations with respect to the latter.

### 2.2. The Hamiltonian Coupled-Mode Theory

In the context of the HCMT, the exact representation of the unknown potential in terms of the free-surface elevation $\eta$ and the modal amplitudes $\varphi_n$, $n \geq -2$, given by Eq. (2), is introduced into the variational equation (6), and the variations are performed with respect to the new independent functional variables $\eta$ and $\varphi_n$. The new Euler-Lagrange equations lead, after an extensive analytical treatment presented in Papoutsellis and Athanassoulis (2017) and Papoutsellis et al. (2018), to the following two Hamiltonian evolution equations with respect to $\eta(x,t)$ and $\psi(x,t) = \Phi(x, z=\eta(x,t), t)$:

$$\partial_t \eta = -(\partial_x \eta) \cdot (\partial_x \psi) + \left((\partial_x \eta)^2 + 1\right)\left(h_0^{-1}\varphi_{-2} + \mu_0 \psi\right), \tag{7a}$$

$$\partial_t \psi = -g\eta - \frac{1}{2}(\partial_x \psi)^2 + \frac{1}{2}\left((\partial_x \eta)^2 + 1\right)\left(h_0^{-1}\varphi_{-2} + \mu_0 \psi\right)^2, \tag{7b}$$

and the following system of horizontal partial differential equations with respect to the modal amplitudes $\varphi_n$, $n \geq -2$:

$$\sum_{n=-2}^{\infty}\left(A_{m,n}\partial_x^2 + B_{m,n}\partial_x + C_{m,n}\right)\varphi_n = 0, \qquad x \in X, \quad m \geq -2, \tag{8a}$$

$$\sum_{n=-2}^{\infty}\varphi_n = \psi, \qquad x \in X, \tag{8b}$$

supplemented by the lateral boundary conditions

$$\sum_{n=-2}^{\infty} A_{m,n}\partial_x \varphi_n + \frac{1}{2}B_{m,n}\varphi_n = 0, \quad x = a, b, \quad m \geq -2. \tag{8c}$$

The coefficients $A_{m,n}$, $B_{m,n}$ and $C_{m,n}$, appearing in the left-hand side of Eqs. (8a) and (8c), are expressed as vertical integrals of the basis functions $Z_n = Z_n(z; \eta, h)$, and are given by the formulae

$$A_{m,n} = \int_{-h}^{\eta} Z_n Z_m \, dz, \tag{9a}$$

$$B_{m,n} = 2\int_{-h}^{\eta}(\partial_x Z_n) Z_m \, dz + \partial_x h \left[Z_n Z_m\right]_{z=-h}, \tag{9b}$$

$$C_{m,n} = \int_{-h}^{\eta}\left(\partial_x^2 Z_n + \partial_z^2 Z_n\right) Z_m \, dz + \left(\partial_x h, 1\right) \cdot \left[\begin{pmatrix}\partial_x Z_n \\ \partial_z Z_n\end{pmatrix} Z_m\right]_{z=-h}. \tag{9c}$$

The following remarks, explaining the structure of the above equations, are in order here.

The evolution equations (7) are not closed with respect to $\eta(x,t)$ and $\psi(x,t)$ since they contain the modal amplitude $\varphi_{-2}$. The latter is obtained by solving the system of equations (8). Since the coefficients of this system are dependent only on $\eta$ and $h$ (see Eqs. (9)), and its excitation is $\psi$ (see Eq. (8b)), we can write

$$\varphi_{-2} = \mathcal{F}[\eta, h]\psi. \tag{10}$$

Eq. (10) reveals that $\varphi_{-2}$ can be considered as a linear, nonlocal operator on $\psi$, also dependent (nonlinearly, yet explicitly) on the boundary functions $\eta$ and $h$. The operator $\mathcal{F}[\eta, h]\psi$ is similar in nature to the classical Dirichlet-to-Neumann operator $\mathcal{G}[\eta, h]\psi$, as defined in Craig and Sulem (1993). In fact, in Athanassoulis and Papoutsellis (2017), the following identity has been proved,

$$\mathcal{G}[\eta, h]\psi = -\partial_x \eta \, \partial_x \psi + \left((\partial_x \eta)^2 + 1\right)\left(\mathcal{F}[\eta, h]\psi/h_0 + \mu_0 \psi\right)^2, \tag{11}$$

permitting us to show that the evolution Eqs. (9) are equivalent to the classical Hamiltonian formulation, as presented by Craig and Sulem (1993). The main difference between the two equivalent formulations lies in the way of treating the nonlocal operators $\mathcal{G}[\eta,h]\psi$ and $\varphi_{-2} = \mathcal{F}[\eta,h]\psi$. The former is usually expanded in a functional Fréchet-Taylor series, which is convergent only for mildly nonlinear waves and mildly non flat bathymetry. The latter is calculated by solving the system (8), which gives accurate results even for strongly nonlinear and very steep (smooth) bathymetries. See Athanassoulis and Papoutsellis (2017) for a detailed discussion on this issue.

A central point for the efficient implementation of the HCMT, Eqs. (7) and (8), is the accurate and fast calculation of the coefficients $A_{m,n}$, $B_{m,n}$, $C_{m,n}$, which, in the numerical procedure, have to be evaluated at each $(x,t)$. Since the vertical basis functions $Z_n$ are explicitly given by means of simple functions, all coefficients $A_{m,n}$, $B_{m,n}$, $C_{m,n}$ can be (and have been) expressed in closed forms; see Papoutsellis et al. (2018), Sec. 4. The analytic expressions of these coefficients greatly accelerate the numerical solution of the substrate problem (8).

Having solved the evolution equations (7), the calculation of any field quantities (pressure, velocity, acceleration) throughout the whole domain is almost *costless*, since it can be performed by simple manipulation of the rapidly convergent series (6). For example, to calculate the hydrodynamic force $F_d(t)$ exerted on the wall, we have to compute the integral

$$F_d(t) = \int_{-h_0}^{\eta(x_w,t)} p_d(x_w,z,t) dz = \int_{-h_0}^{\eta(x_w,t)} \left[ \partial_t \Phi + \frac{1}{2}(\nabla \Phi)^2 \right]_{x=x_w} dz, \qquad (12)$$

where $p_d(x_w,z,t)$ is the hydrodynamic pressure at the position of the wall $x = x_w = b$. The differentiations appearing in the integrand of the last term of Eq. (12) can be performed term-wise on the series expansion (2), leading to (recall here that $\varphi_n = \varphi_n(x,t)$ and $Z_n = Z_n(z;\eta,h)$)

$$\partial_t \Phi = \sum_{n=-2}^{\infty} (\partial_t \varphi_n Z_n + \varphi_n \partial_t Z_n), \quad \partial_t Z_n = \partial_\eta Z_n \partial_t \eta, \qquad (13a)$$

$$\partial_x \Phi = \sum_{n=-2}^{\infty} (\partial_x \varphi_n Z_n + \varphi_n \partial_x Z_n), \quad \partial_x Z_n = \partial_h Z_n \partial_x h + \partial_\eta Z_n \partial_x \eta, \qquad (13b)$$

$$\partial_z \Phi = \sum_{n=-2}^{\infty} \varphi_n \partial_z Z_n. \qquad (13c)$$

Then, the vertical integrals of the $Z_n$−functions and their derivatives, involved in the last term of Eq. (12), can be computed analytically, exploiting the results of Sec. 3 and Sec. 4 of Papoutsellis et al. (2018). Thus, the formula for the hydrodynamic force takes the form

$$F_d(t) = A_1 \partial_t \eta(x_w,t) + A_2 \partial_x \eta(x_w,t) + \\ + \sum_{n=-2}^{\infty} \left( B_n \partial_t \varphi_n(x_w,t) + C_n \partial_x \varphi_n(x_w,t) \right), \qquad (14)$$

where $A_1, A_2, B_n, C_n$ are complicated coefficients, calculated in closed forms. In this way, no further numerical differentiation or integration is performed in computing the force, its accuracy being the same as the accuracy of the free-surface calculation. Recall also that the term-wise differentiated series (13) are rapidly convergent as well, thus, the truncation error is the same as that of the solution procedure.

## 3. A concise description of the numerical implementation

In this section we briefly discuss the numerical implementation of the HCMT, for the 2D case. For a detailed description, the interested reader is referred to Papoutsellis et al. (2018) and Papathanasiou et al. (2019a). These papers also contain extensive numerical investigation and comparisons with experiments and other solution methods, validating the high accuracy, rapid convergence and overall efficiency of the HCMT.

### 3.1. Numerical solution of the kinematical substrate problem

To march in time the two evolution Eqs. (7), we first need to compute the modal amplitude $\varphi_{-2} = \mathcal{F}[\eta, h]\psi$, which is included in their right-hand side. For that purpose, the series expansion (2) is truncated at $n = M$, keeping a finite number of $N_{tot} = M + 3$ modes. The $N_{tot}$ unknown modal amplitudes $\{\varphi_n\}_{n=-2}^{M}$ are computed by numerically solving the system of the first $N_{tot} - 1$ differential equations (8a), with the involved series truncated at $n = M$ as well, together with the algebraic condition (8b), also truncated at $n = M$. In this way, a square linear system is obtained, which is solved numerically by using fourth-order central finite differences, on a uniform grid $\{x_i\}_{i=1}^{N_X}$ of spacing $\Delta x$ spanning $X$. In particular, the first and second $x$-derivatives of each $\varphi_n$, at each point $x_i$, are calculated by the fourth-order central-differences equations

$$(\partial_x u)^i = \frac{1}{12\Delta x}(u^{i-2} - 8u^{i-1} + 8u^{i+1} - u^{i+2}), \tag{15a}$$

$$(\partial_x^2 u)^i = \frac{1}{12\Delta x^2}(-u^{i-2} + 16u^{i-1} - 30u^i + 16u^{i+1} - u^{i+2}), \tag{15b}$$

where $u^i \equiv u(x_i)$. At the end points, $i = 1, N_X$, (resp., the adjacent points, $i = 2, N_X - 1$), of the computational grid, instead of Eqs. (15), one-sided (resp., asymmetric) fourth-order finite differences are used. Substituting Eqs. (15), along with their one-sided (asymmetric) counterparts into the truncated Eqs. (8), leads to a $(M+3)N_X \times (M+3)N_X$ linear algebraic system, whose explicit form can be found in Appendix D of Papoutsellis et al. (2018). This system is numerically solved by using the LU decomposition.

### 3.2 Numerical solution of the Hamiltonian evolution equations

Having established a procedure for the numerical calculation of $\varphi_{-2}^i = (\mathcal{F}[\eta, h]\psi)^i$ at every local position $x_i$ (and every time instant), the Hamiltonian evolution Eqs. (7) are rendered closed, so that their time marching, via a time-integration scheme, is possible. Using again Eqs. (15) and their one-sided (asymmetric) versions, for the approximation of the spatial derivatives contained in Eqs. (7), the latter turn into the following, semi-discrete, evolution system

$$(\partial_t \eta)^i = -(\partial_x \eta)^i (\partial_x \psi)^i + \left([(\partial_x \eta)^i]^2 + 1\right)\left(h_0^{-1}(\mathcal{F}[\eta, h]\psi)^i + \mu_0 \psi^i\right), \tag{16a}$$

$$(\partial_t \psi)^i = -g\eta^i - \frac{1}{2}[(\partial_x \psi)^i]^2 + \frac{1}{2}\left([(\partial_x \eta)^i]^2 + 1\right)\left(h_0^{-1}(\mathcal{F}[\eta, h]\psi)^i + \mu_0 \psi^i\right)^2, \tag{16b}$$

where $(f)^i = f(x_i, t)$, $i = 1, 2, ..., N_X$, for $f = \eta, \psi$. This system is treated by using the classical, fourth-order Runge-Kutta method, choosing a uniform temporal grid $t^n = n\Delta t$, $n = 0, 1, 2, ...$, $\Delta t$ being the uniform time step in use ([3]). The above explicit scheme has been extensively studied and validated against several experiments, concerning water-wave problems with solitary waves, showing very good conservation properties, which are preserved even in long time simulations (Papoutsellis et al. 2018). Let it be noted that we do not need (and do not use) any filtering procedure to obtain accurate results even in long time simulations.

The required initial condition, for the problems studied in this paper, is the flow corresponding to a solitary wave located well before the first trench (and thus non interacting with it). As already mentioned, this nonlinear initial flow is obtained by means of a highly accurate, iterative solver, constructed by Clamond and Dutykh (2013), and based on the full Euler equations. A short overview of this method is presented in Appendix A.

## 4. Propagation of a solitary wave over some typical bathymetries. Validation and limitations of the present method

In this section, we present some comparisons of numerical results obtained via the HCMT for the transformation of a solitary wave by varying bathymetries with available experimental results and results by other nonlinear solvers. This process allows us to validate various features of the HCMT, and identify cases where the assumption of potential flow is inadequate, as viscous effects become important. Our main concern is the study of the interaction of a solitary wave with trench(es). However, we were able to find only one already studied relevant configuration, for which experimental as well as numerical results were available (Chang et al. 2011; Chu et al. 2015). Thus, we enlarge the scope of comparisons, considering also shoaling-type bathymetries, for which several previous studies can be found. Let it be noted that, generally, these cases are more demanding than the ones with trenches, since the lesser the depth the more prominent the viscous effects become.

### 4.1. Propagation over a shelf

As a first case study, we consider the propagation of a solitary wave over a shelf, using the configuration shown in Fig. 1. This problem has been treated in the works of Madsen and Mei (1969), providing both experimental results and numerical ones based on approximate long-wave equations, and Li et al. (2012), who implement a Smoothed Particle Hydrodynamics (SPH) model that solves the weakly compressible Navier–Stokes equations. The incident solitary wave initially propagates over a horizontal seabed of depth $h_0$ and, then, after climbing up a mildly sloping seabed, continues to propagate over a shelf of depth $h_1 < h_0$. During its evolution, the free-surface elevation is measured at the gauges $g_1$, $g_2$, $g_3$ and $g_4$.

The parameters required for the simulation of the problem via the HCMT are determined as follows. The horizontal computational domain is $X = [-5m, 8m]$, with the initial solitary wave centered at $x = 0$ and the sloping seabed extending from $x = 2m$ to $x = 2.762m$. For $x \in [-5m, 2m]$, the water depth is $h_0 = 0.0762m$, while, for $x \in [2.762m, 8m]$, the depth becomes $h_1 = h_0/2 = 0.0381m$. In between, it is determined by requiring the slope to be 1:20. The gauges are placed at positions $x_{g_1} = 1.5936m$, $x_{g_2} = 2.762m$, $x_{g_3} = 3.651m$ and $x_{g_4} = 4.5146m$. The amplitude of the incident solitary wave is $a/h_0 = 0.12$ and its velocity $c = 0.9145 m/s$. The numerical solution via the HCMT is realized using $N_{tot} = 8$ modes ($\varphi_{-2}, \varphi_{-1}, \varphi_0, \varphi_1, \cdots, \varphi_5$), spatial discretization $\Delta x/h_0 = 0.08$ and temporal discretization $\Delta t$ such that the Courant number, in terms of the velocity $c$ of the initial solitary wave, takes the (constant) value $C = 0.5$.

---

([3]) Another time-integration scheme, exploiting the analytic solution of a linear part of the evolution equations (exponential integrators), is currently under development (Papathanasiou, Papoutsellis and Athanassoulis 2019b).

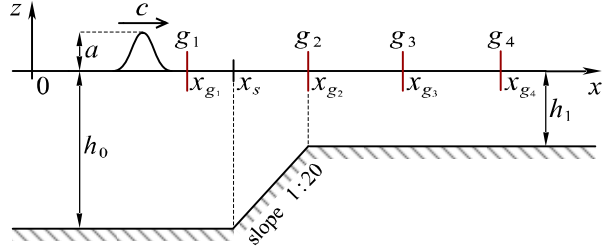

**Fig. 1** Configuration of the propagation of a solitary wave over a shelf: $h_0 = 0.0762\,m$, $a/h_0 = 0.12$, $h_1 = 0.0381\,m$, $x_{g_1} = 1.5936\,m$, $x_s = 2\,m$, $x_{g_2} = 2.762\,m$, $x_{g_3} = 3.651\,m$, $x_{g_4} = 4.5146\,m$

As demonstrated in Fig. 2, our computations are in good agreement with the referenced experimental data and almost identical with the numerical results obtained with the SPH approach, up to the gauge $g_2$. Past that point, a deviation from the experimental measurements and, to a lesser extent, from the SPH results occurs. Specifically, our results are characterized by higher peaks and an advancement of the wave front. To quantify this discrepancy, we mention that, at the gauge $g_3$, the highest-peak overestimation with respect to the experimental data is about 14.1%, with a time advancement of about 1.6%. At the gauge $g_4$, the respective percentages are shaped as 36.4%, for the highest peak, and 1.6%, for the time advancement. As a point of reference, we should also note that the highest-peak overestimation of the SPH results compared to the experimental measurements is about 8.1% and 22.1% at the gauges $g_3$ and $g_4$, respectively. This behavior can generally be attributed to viscous effects (especially, bottom friction) that become more significant in the shallower region, where the two gauges $g_3, g_4$ are located. Furthermore, it should be noted that in this experiment the physical depth is very small ($h_1 = 3.81\,cm$), rendering the depth-based Reynolds number very small as well, and the bottom friction significant for the flow, for a large part of the water column. We further elaborate on this point in Sec. 4.3. Note finally that, in all cases, our numerical results concerning the free-surface elevation are more consistent with the experimental ones than the numerical results of Madsen and Mei (1969), obtained in terms of long-wave equations. The alignment of the main peaks between their experimental and numerical results seems to be artificial, since the viscous effects must decrease the propagation speed. This is also supported by the fact that even the SPH solution does not provide a peak fully aligned with the experimental one.

### 4.2. Propagation over a step

Next, we study the propagation of a solitary wave over a step change in bathymetry. To make this bathymetry compatible with our implementation, the depth discontinuity has been modelled as a smooth, very steep, continuous change, by using the following tanh-function representation

$$h(x) = h_0 + \frac{h_1 - h_0}{2}\left[\tanh(sx) + 1\right]. \tag{17}$$

Referring to Fig. 3, the incident solitary wave propagates, for a while, over a seabed of depth $h_0$, before encountering the step. Then, during its interaction with the step, a wave transformation process occurs, and finally the resulting wave system evolves over a shallower region, of depth $h_1$. While the wave propagation/transformation takes place, the elevation of the free surface is measured at four gauge locations, i.e. gauges $g_1$, $g_2$, $g_3$ and $g_4$.

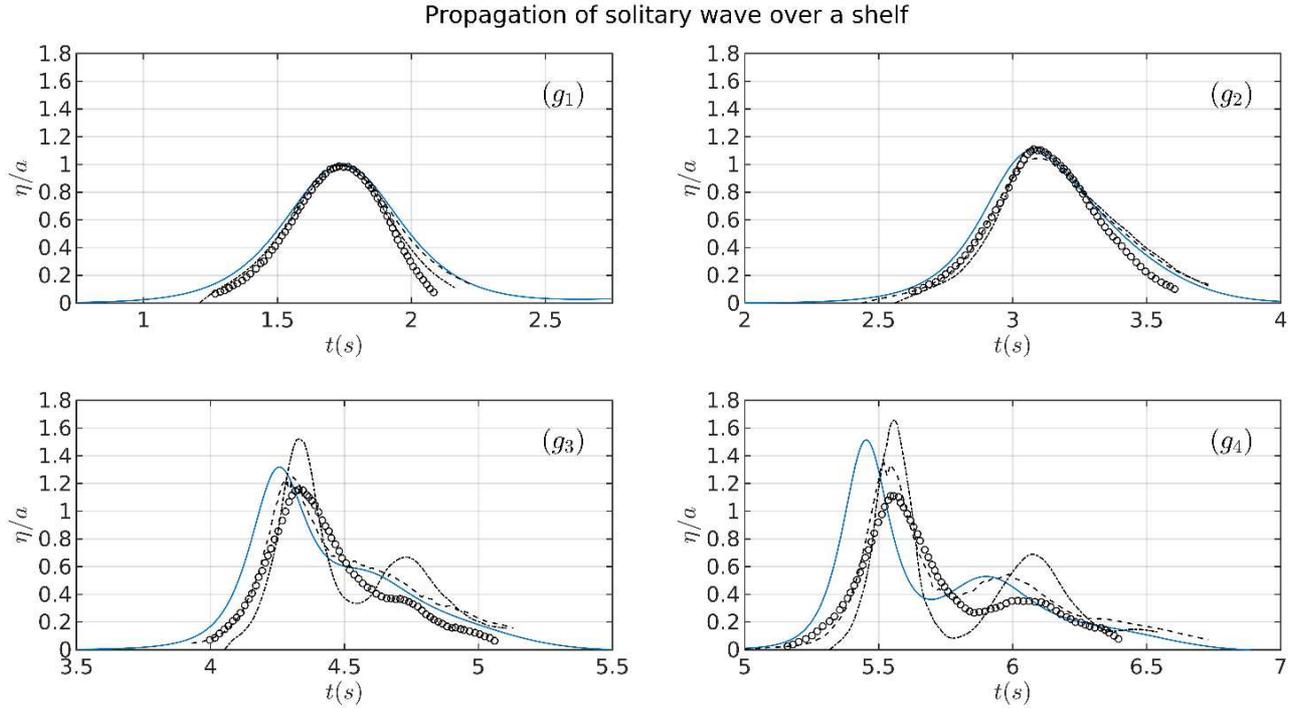

**Fig. 2** Comparison of the normalized free-surface elevation history at the gauges $g_1$, $g_2$, $g_3$ and $g_4$, between experimental data (circles) and numerical results (dash-dot line) from Madsen and Mei (1969), numerical results by means of the SPH method from Li et al. (2012) (dashed line) and HCMT numerical results (solid blue line)

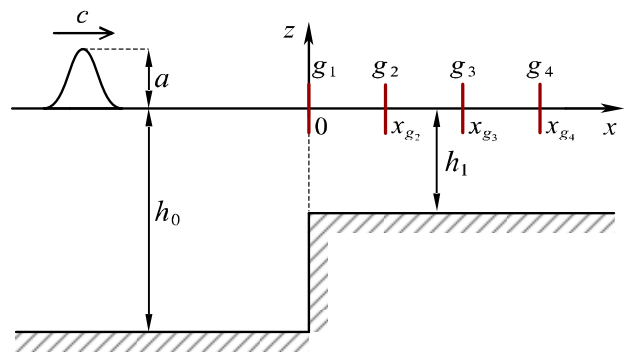

**Fig. 3** Configuration of the propagation of a solitary wave over a shelf: $h_0 = 0.2\,m$, $a/h_0 = 0.182$, $h_1/h_0 = 0.5$, $x_{g_1}/h_0 = 0$, $x_{g_2}/h_0 = 15$, $x_{g_3}/h_0 = 30$ and $x_{g_4}/h_0 = 45$

To be able to compare our numerical results with others', the simulation is performed with the configuration used by Seabra-Santos et al. (1987) and P. L.-F. Liu and Cheng (2001). The former presented experimental as well as numerical results (with long-wave equations including curvature effects), whereas the latter treated the problem by means of a Reynolds Averaged Navier-Stokes (RANS) model coupled with $k-\varepsilon$ turbulence equations. The geometry of the domain is defined by $h_0 = 0.2\,m$ and $h_1/h_0 = 0.5$, and the amplitude of the initial solitary wave is such that $a/h_0 = 0.182$, leading to a velocity $c = 1.5213\,m/s$. The horizontal domain extends from $x/h_0 = -60$ to $x/h_0 = 60$, with the bathymetry change happening at $x = 0$. The gauges are placed at positions $x_{g_1}/h_0 = 0$, $x_{g_2}/h_0 = 15$, $x_{g_3}/h_0 = 30$ and $x_{g_4}/h_0 = 45$.

Our results, via the HCMT, are shown in Fig. 4. Again, a very good agreement with the referenced works takes place at the first two gauges, while at gauges $g_3$ and $g_4$, a deviation from the experimental results occurs, due to reasons already mentioned in Sec. 4.1 (viscous effects, development of vorticity at the edge). It is interesting, and somewhat surprising, that the numerical results obtained via the RANS model of P. L.-F. Liu and Cheng (2001) also exhibit a significant deviation from the experimental ones, being closer to ours (than to the experimental values). To be more precise, the HCMT highest-peak overestimation, compared to the experimental data, is 18.3% and 20.1% at the gauges $g_3$ and $g_4$, respectively, whereas the time advancement at each gauge is about 3% and 2%. The corresponding deviations of the RANS results from the experimental ones, for the highest-peak overestimation and the time advancement are: 12.1% and 2.2% at the gauge $g_3$, and 12.4% and 1.5% at the gauge $g_4$. Two conclusions can be drawn from these findings. First, the viscous effects, emerging in the vicinity of the step, should be severe and, second, regarding the free-surface elevation, the fully nonlinear potential theory gives satisfactory results, well compared with those obtained via CFD methods, while being much lighter computationally. As for the numerical results of Seabra-Santos et al. (1987), a better performance regarding the main wave front at the gauges $g_3$ and $g_4$ is evident, which is degraded subsequently, leading to worse predictions at the second peak.

### 4.3. Comparison and assessment of the findings of Sec. 4.1 and 4.2

Comparing the findings of Sec. 4.1 and 4.2, we are facing a striking and somewhat counter-intuitive result; the percentage discrepancy between our numerical results and the experimental ones is generally larger in the case of a shelf, where the transition to the half depth is mild, in comparison with the corresponding percentage discrepancy in the case of a step, where the respective transition to the half depth is much steeper. In other words, it appears that the HCMT solution is able to follow the experimental results more efficiently in the more demanding case of a step-change in bathymetry (detailed figures of the peak overestimation and time advancement for the two peaks of a solitary wave transformed by the shelf and the step, considered in Sec. 4.1 and 4.2, are presented in Appendix B). This observation becomes even more peculiar if we consider that the amplitude of the solitary wave in the second case is higher than in the first.

As was briefly mentioned in Sec. 4.1, this situation should be attributed to the fact that the physical depth in the case of the shelf experiment is smaller than the physical depth in the case of the step experiment. A physically sound quantification of this difference can be made by considering the appropriate (depth-based) Reynolds number of the corresponding flows: $Re = h\sqrt{gh}/\nu$, where $\nu$ is the kinematic viscosity of the water. Denoting by $h_{\text{shelf}} = 3.81\,cm$ and $h_{\text{step}} = 10\,cm$ the depth in the shallow region of the shelf and the step experiment, respectively, and by $Re_{\text{shelf}}$ and $Re_{\text{step}}$ the corresponding Reynolds numbers, we obtain

$$\frac{\mathrm{Re}_{\mathrm{shelf}}}{\mathrm{Re}_{\mathrm{step}}} = \left(\frac{h_{\mathrm{shelf}}}{h_{\mathrm{step}}}\right)^{3/2} \implies \mathrm{Re}_{\mathrm{step}} \approx 4.25\,\mathrm{Re}_{\mathrm{shelf}}.$$

It is well documented (see, e.g., Tang et al. (1990)) that the smaller the Reynolds the larger the impact of the viscous effects on the solitary wave, which affects its propagation characteristics, causing the reduction of its amplitude and velocity. That is, although in the step experiment the change in bathymetry is much steeper, the flow is governed by a Reynolds number more than four times greater, which results in milder viscous effects on the wave flow, in comparison with the case of the shelf experiment. That explains why the HCMT, being a potential theory, performs better in the case of Sec. 4.2.

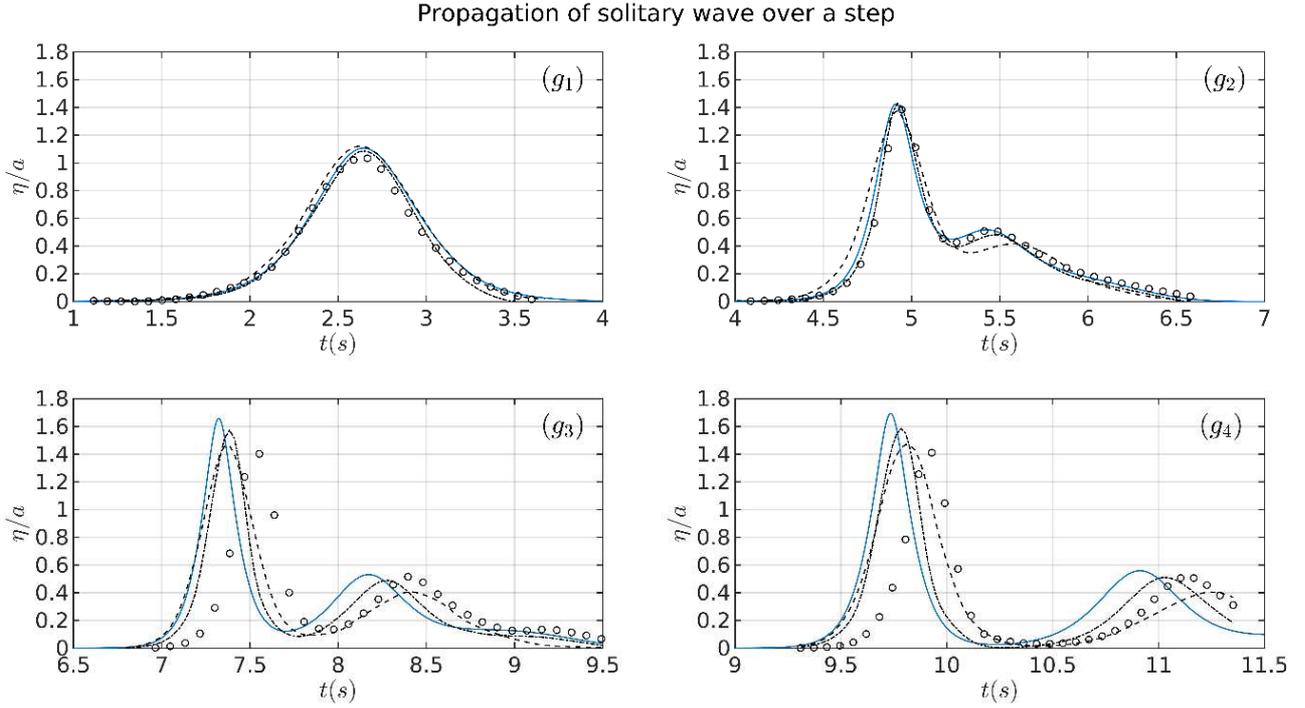

**Fig. 4** Comparison of the normalized free-surface elevation history at the gauges $g_1$, $g_2$, $g_3$ and $g_4$, between experimental data (circles) and numerical results from Seabra-Santos et al. (1987) (dashed line), numerical results by means of a RANS model from P. L.-F. Liu and Cheng (2001) (dash-dot line) and HCMT results (solid blue line)

### 4.4. Propagation over a trench

Having examined cases of shoaling bathymetries, we now proceed with the investigation of the interaction of a solitary wave with a trench, which is the most interesting and most relevant case in regard to our applications([4]). In this case, as illustrated in Fig. 5, an incident solitary wave, propagating initially over a uniform seabed of depth $h_0$, passes over a trench of length $l_{tr}$, where the local depth is $h_0 + h_{tr}$. Such configurations have been studied by Chang et al. (2011), both experimentally and numerically, using the Finite-Analytic (FA) method to solve the 2D Navier-Stokes equations in stream function and vorticity formulation. The free-surface boundary conditions are treated via a Finite Difference scheme.

---

([4]) Again, to make the bathymetry with a rectangular trench compatible with the smooth-bathymetry requirement of the HCMT, the trench has been modelled as a smooth continuous change, by using a combination of two tanh-functions. More details are given in Sec. 5.

During that wave motion, the free-surface elevation is measured at the gauges $g_1$, $g_2$ and $g_3$, located before, at the end, and after the trench.

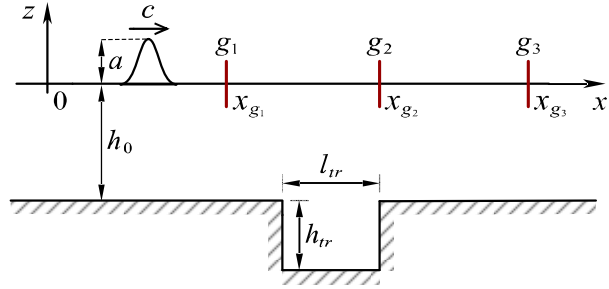

**Fig. 5** Configuration of the propagation of a solitary wave over a trench: $h_0 = 0.0762\,m$, $a/h_0 = 0.41, 0.42$, $l_{tr}/h_0 = 5, 3.5$, $h_{tr}/h_0 = 1$, $x_{g_1}/h_0 = -10.6$, $x_{g_2}/h_0 = 5$, $x_{g_3}/h_0 = 19$

For validation purposes, we solve this problem by using the HCMT, adopting the same setting as Chang et al. (2011), who perform the experiment twice, for two different values of the trench's length, $l_{tr}/h_0 = 5$ and $l_{tr}/h_0 = 3.5$. In both cases, the physical depth is $h_0 = 0.0762\,m$, the trench height is $h_{tr}/h_0 = 1$, and the trench begins at $x/h_0 = 0$. The amplitudes of the initial solitary waves are $a/h_0 = 0.41$ and $a/h_0 = 0.42$, respectively, resulting in propagation velocities $c = 1.0218\,m/s$ and $c = 1.0251\,m/s$. The positions of the gauges $g_1$, $g_2$, $g_3$ are, respectively, set at $x_{g_1}/h_0 = -10.6$, $x_{g_2}/h_0 = 5$ and $x_{g_3}/h_0 = 19$. As a result, the horizontal computational domain is chosen to be $X = [-40 h_0, 40 h_0]$.

As shown in Fig. 6, our results are practically identical with the numerical ones of Chang et al. (2011), exhibiting a small discrepancy with respect to the experimental results, especially just after the wave front passes through the gauges. That can be justified by the fact that, in contrast to the previous cases, the vortical effects generated at the region of the trench weaken significantly as they approach the free surface, remaining essentially restricted near the trench and in a thin layer near the bottom after the trench; see e.g. Fig. 7 – 10 of Chang et al. (2011). Therefore, we can conclude that the impact of the vortical flow on the evolution of the free surface and the main wave flow is small.

## 5. Propagation of a solitary wave over two trenches and reflection at a vertical wall

In this section, we focus on the study of the propagation of a non-breaking solitary wave over two tandem trenches and its subsequent reflection at a vertical wall. Special attention is paid to the run-up and the force exerted on the wall. The case of one trench is also considered.

The geometric configuration of the cases studied herein is shown in Fig. 7. The main parameters are: the central points of the two trenches, $x_1$ and $x_2$, the position of the wall, $x_w = b$, and the height and length of each trench, $h_{tr}$, $l_{tr}$, respectively. To systematize the investigation, we set $h_0 = 1\,m$, and fix the centers at the positions $x_1/h_0 = 0$ and $x_2/h_0 = 56$, respectively, whereas the wall is located at $x_w/h_0 = 136$. Three cases are considered regarding the trenches' height, $h_{tr}/h_0 = 0.5, 1, 2$, while the effect of their length is investigated for the values $l_{tr}/h_0 = 4, 8, 12, 16, 24, 36$. The amplitude of the (nonbreaking) incident solitary wave is chosen as $a/h_0 = 0.2, 0.3, 0.4$. To better understand the effect of the two tandem trenches, and for comparison reasons, the same numerical experiments are repeated with only one trench, that at position $x_2$.

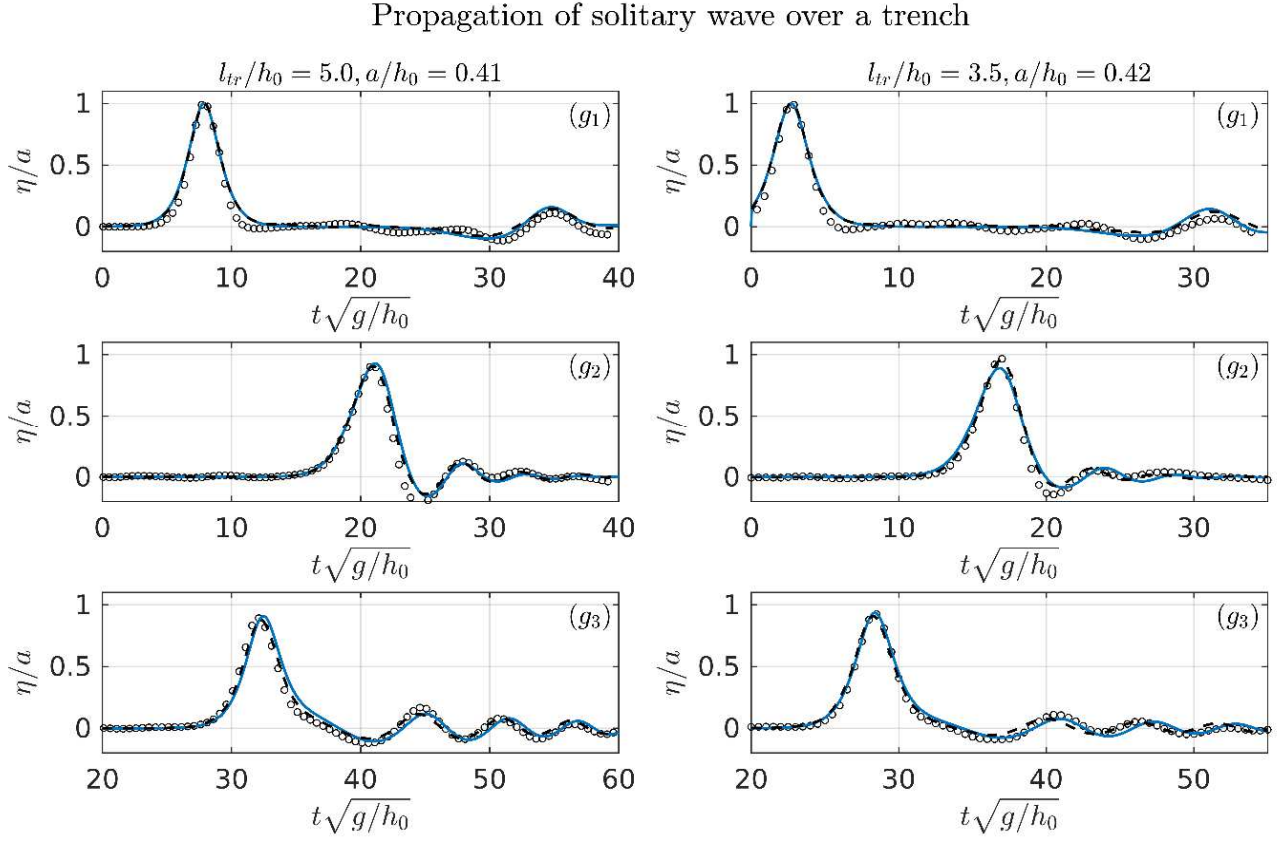

**Fig. 6** Comparison of the free-surface elevation history at the gauges $g_1$, $g_2$ and $g_3$, between experimental data (circles) and numerical results (dashed line) by means of the FA method for the 2D Navier-Stokes equations from (Chang et al. 2011) and HCMT results (solid blue line)

The presence of the rectangular trenches introduces a discontinuity at the bottom profile, which, as mentioned earlier, is not compatible with the HCMT. To overcome that difficulty, the sides of the trenches are smoothed out by using the tanh (hyperbolic tangent) function. Specifically, when two trenches are involved, the bathymetry used in the numerical calculations is defined by

$$h(x) = h_0 + \frac{h_{tr}}{4} \{ \tanh[s(x-\xi_{1-})]+1 \} \{ \tanh[s(\xi_{1+}-x)]+1 \} + \\ + \frac{h_{tr}}{4} \{ \tanh[s(x-\xi_{2-})]+1 \} \{ \tanh[s(\xi_{2+}-x)]+1 \}, \quad (18)$$

where

$$\xi_{1\mp} = x_1 \mp l_{tr}/2 \quad \text{and} \quad \xi_{2\mp} = x_2 \mp l_{tr}/2,$$

and $s$ is a constant controlling the steepness of the trench sides. When only the second trench is present (centered at $x_2$), the second term of the right-hand side of Eq. (18) is omitted.

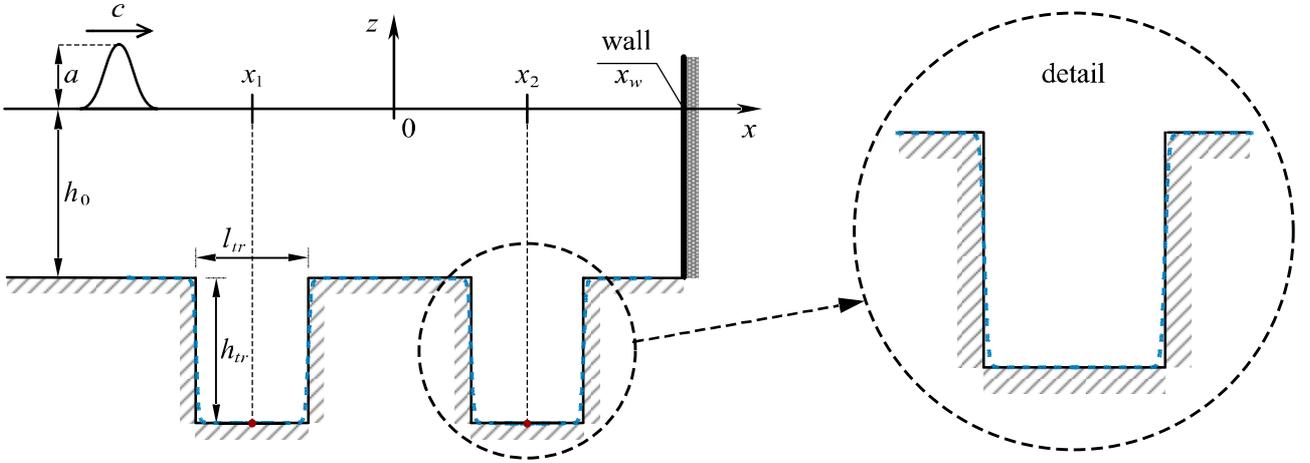

**Fig. 7** Geometric configuration of the studied cases (not scaled). The detail shows the real trench (solid line) versus its smoothed-out version by means of the tanh-function (dashed blue line); Eq. (18)

All the results shown below have been obtained by using the HCMT, Eqs. (7) and (8), Sec. 2, solved as described in Sec. 3. The number of modes used is $N_{tot} = 8$, and the spatial discretization is $\Delta x / h_0 = 0.04$, while the temporal discretization $\Delta t$, is taken so that the Courant number, calculated by using the velocity of the initial solitary wave, is $C = 0.5$.

**5.1. Propagation over the trenches**

We begin our investigation with the study of the free-surface evolution, as the solitary wave propagates over the trenches. As can be seen in Fig. 8, at each trench the wave experiences partial reflection and, further, a dispersive trail emerges behind its front. Apparently, an increase in $h_{tr}$ renders those phenomena more pronounced (compare different subplots in the same row). Concerning the effects of $l_{tr}$, we observe that it strongly affects the dispersive trail, but has a milder impact on the wave reflection (compare different subplots in the same column). The latter can also be confirmed by the snapshots of the free-surface elevation, shown in Fig. 9. The most important result, better observed in Fig. 9, is that the presence of the trenches leads to a significant reduction of the maximum wave amplitude that reaches the wall. Especially for the case of deeper and larger trenches, this reduction can be about 40% of the amplitude of the initial solitary wave; see Fig. 9d. The trench height $h_{tr}$ mainly affects the leading wave, having a weaker effect on the form of the wave trail, as can be seen in Fig. 10.

Secondary effects can be recognized in Figs. 8-10, as well. In particular, as is expected, a minor reflection back to the right direction occurs when the reflected-at-the-second-trench waves reach the first trench again. Moreover, the solitary wave exhibits a slow, fission-like behavior after each trench. That can be better seen in Fig. 9 and 10, and is further supported by the increase in height of the leading wave just after the trenches (see dashed lines in the same figures).

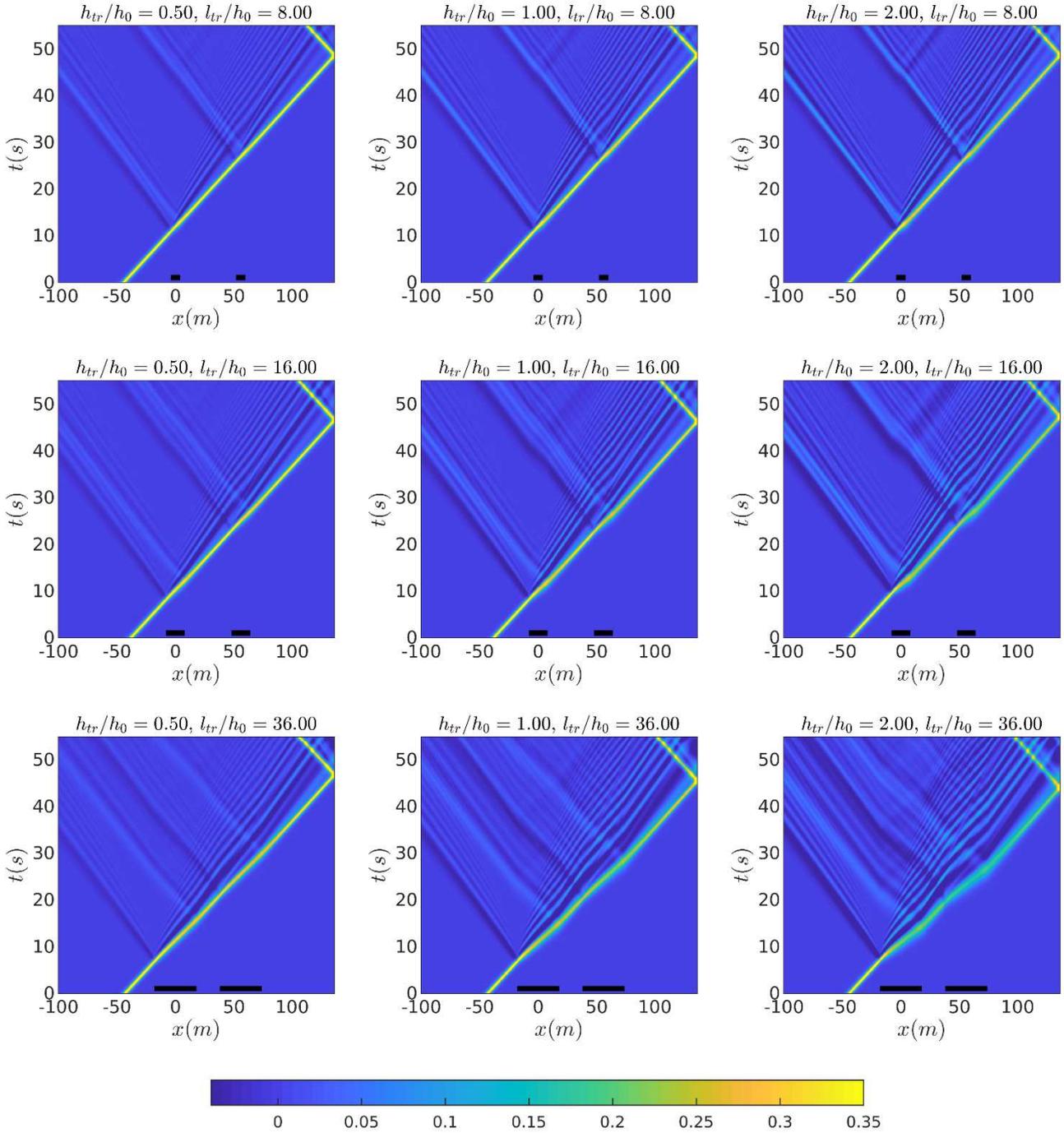

**Fig. 8** Colormaps of the evolution of the free-surface elevation $\eta$ (the vertical axis in each subplot is time), for various configurations of the trenches. The subplots of each row correspond to the same $l_{tr}$ with increasing $h_{tr}$ to the right, while in each column we have the same $h_{tr}$ with increasing $l_{tr}$ from top to bottom. The amplitude of the solitary wave is $a/h_0 = 0.4$. The locations of the trenches are denoted by the small-and-thick black lines at the base of each subplot

The evolution of the free-surface elevation of solitary waves with amplitude $a/h_0 = 0.3$, $0.4$, propagating over two trenches with $h_{tr}/h_0 = 2$ and $l_{tr}/h_0 = 16$, is available in video format. Similar results for the cases $a/h_0 = 0.4$, $l_{tr}/h_0 = 16$ and $h_{tr}/h_0 = 1, 2$ can be viewed here. The evolution of the first five modes ($\varphi_{-2}, \varphi_{-1}, \varphi_0, \varphi_1, \varphi_2$), the free-surface potential $\psi$, and the free-surface elevation $\eta$, for $a/h_0 = 0.4$, $h_{tr}/h_0 = 2$ and $l_{tr}/h_0 = 16$, are presented in a third video. In the last video it can be clearly seen that the bottom mode, $\varphi_{-1}$, becomes important only locally, when the wave passes over the abrupt changes in bathymetry (trench edges). A similar behavior is observed for the evanescent modes, $\varphi_1, \varphi_2$, although their effect is more spread in space and time.

## 5.2. Run-up on the wall

In this subsection, we focus on the quantification of the effect of the trenches on the run-up of the waves reaching the wall. As can be expected, the presence of the trenches has a diminishing effect on the maximum run-up, which is intensified as the height and/or the length of the trenches increases, leading up to 45.7% reduction ([5]) in the cases considered herein.

The nondimensional maximum run-up $\eta_w/a$, for wave amplitudes $a/h_0 = 0.2, 0.3, 0.4$, is shown in Figs. 11-13, respectively, as a function of the nondimensional trench length $l_{tr}/h_0$, and with $h_{tr}/h_0$ as a parameter (recall that $l_{tr}$ is the length of *one* of the two trenches). These figures include both the case of two tandem trenches (solid lines) and the case of a single trench centered at $x_2$ (dashed lines). The case of a flat bottom is conventionally included in the same figure, at $l_{tr} = 0$. Evidently, for all cases, the maximum run-up is a decreasing function of $l_{tr}/h_0$, the decrease being stronger as $h_{tr}/h_0$ increases. The presence of two tandem trenches results in an increase of the run-up reduction by a factor $1.4 – 1.7$, in comparison with the single-trench configuration. Detailed figures of the reduction percentages for $l_{tr}/h_0 = 8, 16, 24, 36$, $h_{tr}/h_0 = 1, 2$, and $a/h_0 = 0.2, 0.3, 0.4$, are shown in Table 1. The maximum reduction percentage, in the presence of two trenches (resp., one trench), is always attained at the maximum values of $l_{tr}/h_0$ and $h_{tr}/h_0$ considered, and reaches 38.4% (resp., 25%) for $a/h_0 = 0.2$, 42.5% (resp., 30.4%) for $a/h_0 = 0.3$ and 45.7% (resp., 34.6%) for $a/h_0 = 0.4$.

Another interesting feature, shown in the same figures, is the weaker impact of the trenches for small values of $l_{tr}/h_0$, which appears as a "curvature change" of the function $\eta_W/a = f(l_{tr}/h_0)$ around the point $l_{tr}/h_0 = 4$, especially for the larger values of $h_{tr}/h_0$. A physical explanation of this feature may be that, for small values of $l_{tr}$ (e.g. $l_{tr}/h_0 = 4$), the effective length of the solitary wave (defined herein as the length of the waveform at $z = 0.05 a$) is considerably larger than the length of the trench and, thus, the wave is only mildly affected.

Obviously, the use of two tandem trenches, instead of a single one, is beneficial for the run-up reduction, no matter the trenches' geometry or the amplitude of the incoming solitary wave. A very interesting point in this context is that, to achieve the same run-up reduction using only one trench, its length must be greater than the total length of the two trenches, for the cases with $l_{tr}/h_0 > 4$ and $h_{tr}/h_0 > 0.5$.

---

([5]) The percentage reduction is calculated by means of the formula $100 (R_{flat} - R_{tr}) / R_{flat} \%$, where $R_{flat}$ is the value of the run-up (or force) in absence of trench(es) (flat seabed), and $R_{tr}$ is the corresponding value when the trench(es) are present.

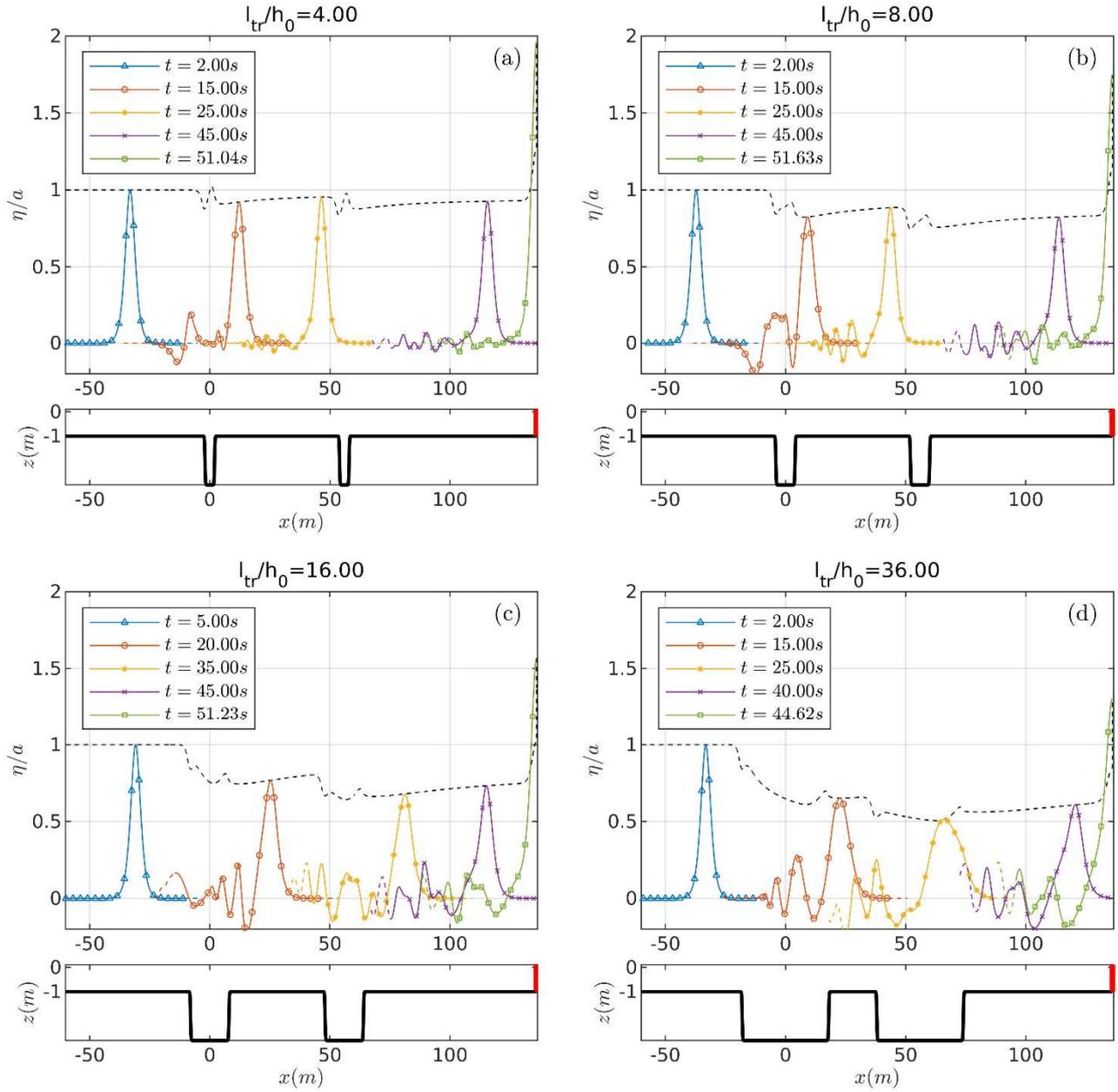

**Fig. 9** Snapshots of the free-surface elevation $\eta/a$ at various time instances, for the cases $l_{tr}/h_0 = 4, 8, 16, 36$ and $h_{tr}/h_0 = 2$, with $a/h_0 = 0.2$. The dashed line denotes the height of the leading wave

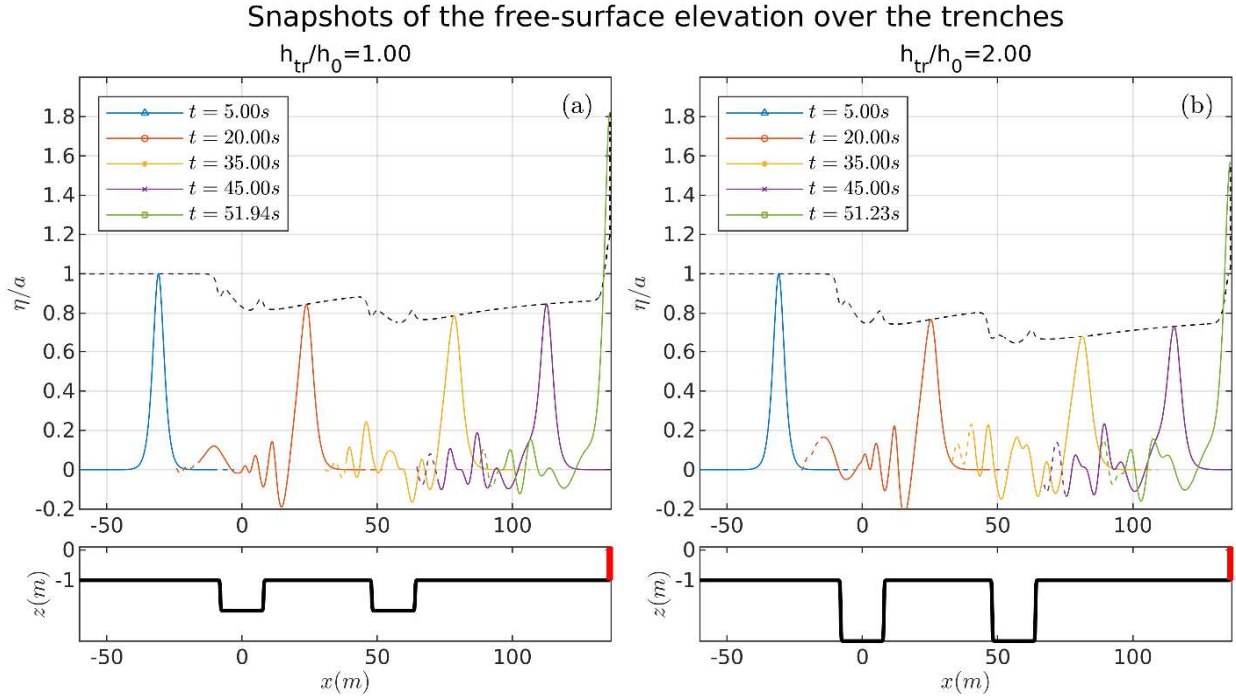

**Fig. 10** Snapshots of the free-surface elevation $\eta/a$ at various time instances, for the cases $h_{tr}/h_0 = 1, 2$ and $l_{tr}/h_0 = 16$, with $a/h_0 = 0.2$. The dashed line denotes the height of the leading wave

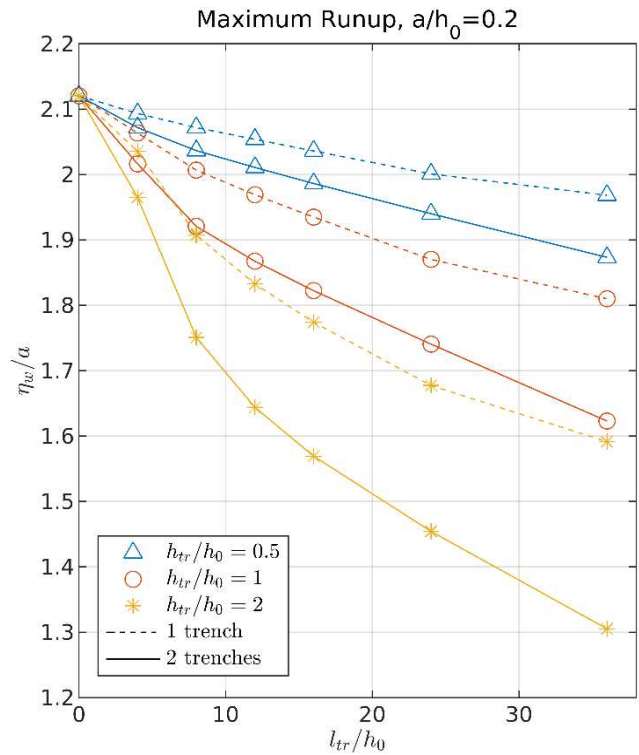

**Fig. 11** The maximum run-up on the vertical wall for the cases of one trench (dashed lines) or two tandem trenches (solid lines), as a function of $l_{tr}/h_0$, for various values of $h_{tr}/h_0$, when $a/h_0 = 0.2$

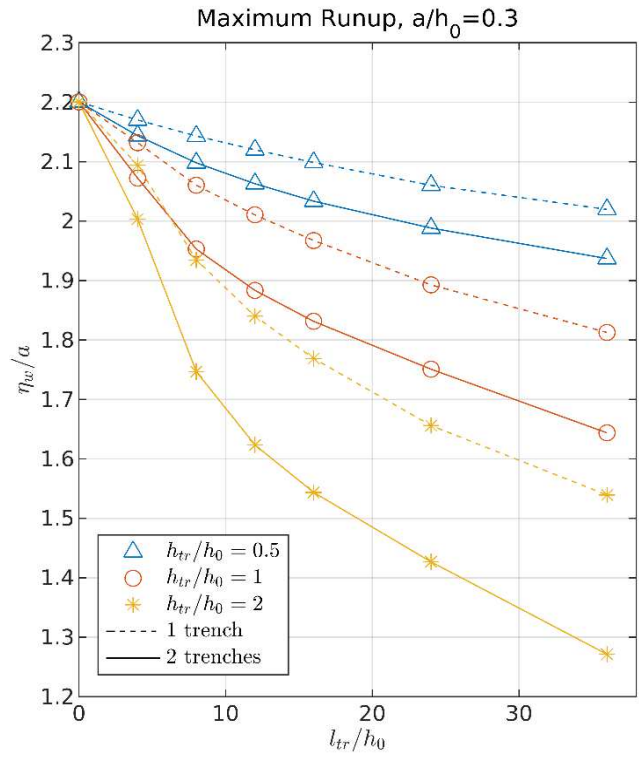

**Fig. 12** The same as in Fig. 11, with $a/h_0 = 0.3$

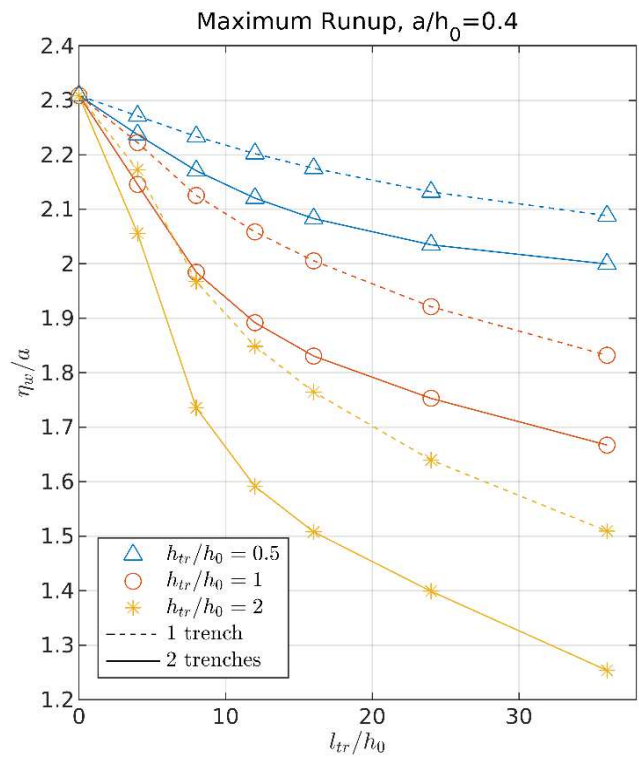

**Fig. 13** The same as in Fig. 11, with $a/h_0 = 0.4$

**Table 1** Percentage reduction of the maximum run-up due to the presence of two tandem trenches (the figures in parenthesis correspond to the one-trench case)

| $h_{tr}/h_0$ | $l_{tr}/h_0$ | | | |
|---|---|---|---|---|
| | 8 | 16 | 24 | 36 |
| $a/h_0 = 0.2$ (effective length / $h_0 = 12.2$) | | | | |
| 1 | 9.2% (5.2%) | 14.2% (8.8%) | 17.9% (11.9%) | 23.5% (14.6%) |
| 2 | 17.4% (10%) | 26% (17.3%) | 31.5% (20.8%) | 38.4% (25%) |
| $a/h_0 = 0.3$ (effective length / $h_0 = 10.3$) | | | | |
| 1 | 11.7% (6.9%) | 17.1% (11%) | 20.6% (14.3%) | 25.6% (17.9%) |
| 2 | 20.9% (12.5%) | 30.1% (20%) | 35.4% (25.1%) | 42.5% (30.4%) |
| $a/h_0 = 0.4$ (effective length / $h_0 = 9.1$) | | | | |
| 1 | 14.1% (8.1%) | 20.8% (13.2%) | 24.1% (16.8%) | 27.8% (20.8%) |
| 2 | 24.8% (14.9%) | 34.7% (23.6%) | 39.4% (29.1%) | 45.7% (34.6%) |

### 5.3. Maximum force exerted on the vertical wall

Now we turn our attention to the force exerted on the wall, focusing on the hydrodynamic part $F_d(t)$, defined in Eq. (12). The maximum value $F_W$ of this force is presented below, in the nondimensional form:

$$\frac{F_W}{F_0} = \frac{\max\left(F_d(t) - F_0\right)}{F_0}, \qquad (19)$$

where $F_0 = 0.5 \rho g h_0^2$ is the mean hydrostatic force. In Figs. 14-16, the quantity $F_W / F_0$ is plotted as a function of $l_{tr}/h_0$, for various values of $h_{tr}/h_0$, and for wave amplitudes $a/h_0 = 0.2, 0.3, 0.4$, respectively. The behavior of the maximum force and its trends with respect to the trench dimensions and the wave amplitude are very similar to those of the run-up. The force percentage reductions due to the presence of the trenches, for various trench dimensions, follow the same pattern as in the case of the run-up, being slightly smaller than the corresponding cases of the latter. Detailed figures of the force reduction percentages, for $l_{tr}/h_0 = 8, 16, 24, 36$, $h_{tr}/h_0 = 1, 2$ and $a/h_0 = 0.2, 0.3, 0.4$, are given in Table 2. The greatest force reduction achieved in the examined cases is about 38%.

### 6. Conclusions

In this work, the fully nonlinear potential Hamiltonian Coupled-Mode Theory was applied to the study of the transformation of a solitary wave by one or two tandem trenches, and to the calculations of the maximum run-up and the maximum force exerted on a vertical wall by the resulting wave system. The nonlinear and dispersive characteristics of the wave flow are fully accounted for, while the vortical flow, which is inevitably developed within and near the edges of the trench(es), is not captured by the present potential theory. Nevertheless, comparisons with experimental results and Navier-Stokes solvers suggest that the vortical effects do not seriously affect the frontal part of the wave flow, justifying the use of the present theory for the calculation of the run-up and the force on the wall. The main result of this investigation is that the presence of two trenches in the bathymetry, in front of the wall, may reduce the run-up from (about) 20% up to 45%, depending on the dimensions of the trench(es), the reduction being greater for higher waves. The corresponding figures for the force are 15% - 38%. These large run-up and force reductions justify the use of trenches as submerged breakwaters even for long nonlinear waves, extending and supplementing the existing analysis in the context of linear wave theories.

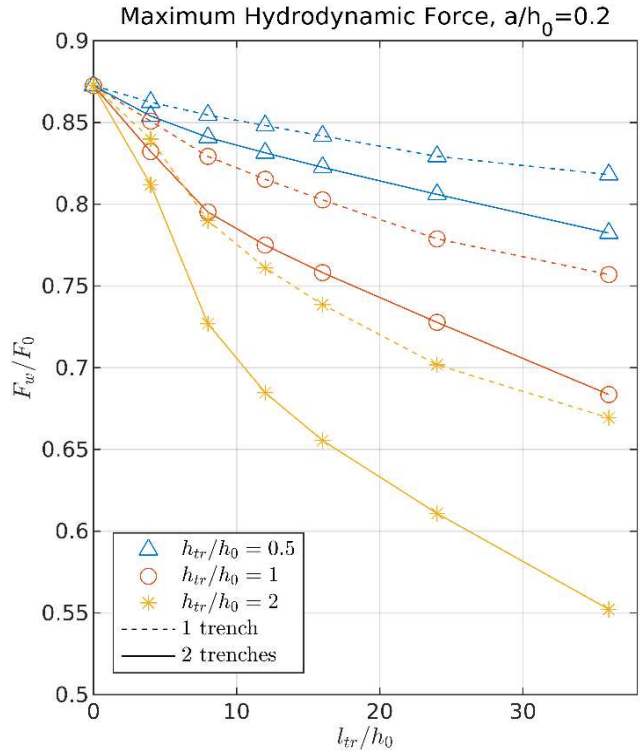

**Fig. 14** The maximum hydrodynamic force on the vertical wall for the cases of one (dashed lines) or two tandem trenches (solid lines), as a function of $l_{tr}/h_0$, for various values of $h_{tr}/h_0$, when $a/h_0 = 0.2$

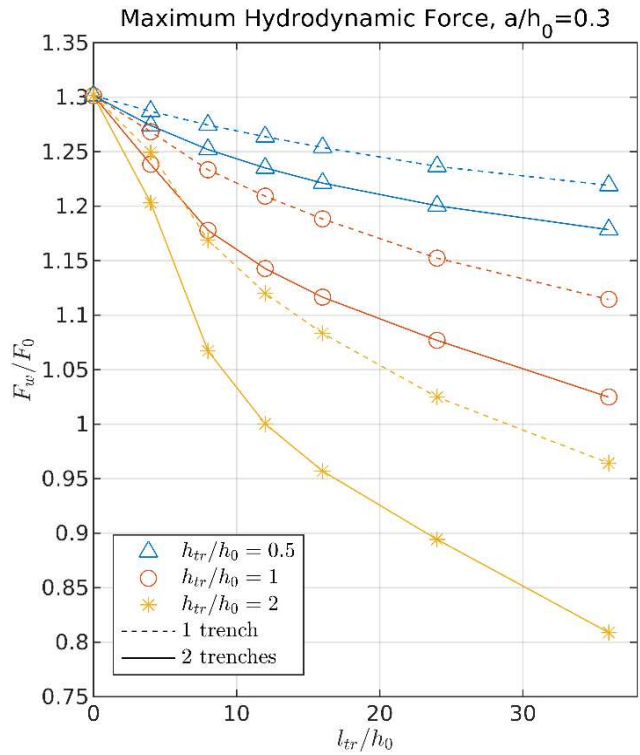

**Fig. 15** The same as in Fig. 14, with $a/h_0 = 0.3$

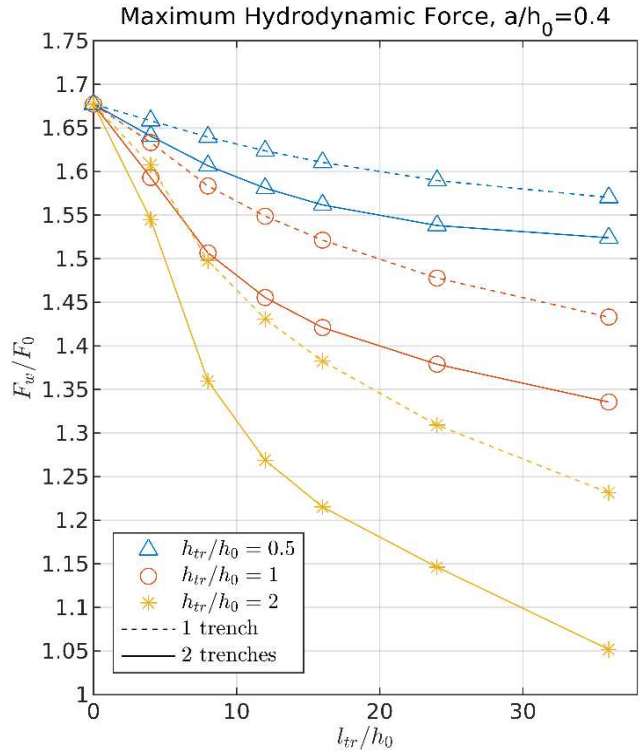

**Fig. 16** The same as in Fig. 14, with $a/h_0 = 0.4$

**Table 2** Percentage reduction of the maximum force due to the presence of two tandem trenches (the figures in parenthesis correspond to the one-trench case)

| $h_{tr}/h_0$ | $l_{tr}/h_0$ | | | |
|---|---|---|---|---|
| | **8** | **16** | **24** | **36** |
| $a/h_0 = 0.2$ (effective length / $h_0 = 12.2$) | | | | |
| 1 | 8.8% (5.1%) | 13.1% (7.9%) | 16.6% (10.7%) | 21.7% (13.2%) |
| 2 | 16.6% (9.4%) | 24.9% (15.4%) | 30.1% (19.5%) | 36.6% (23.3%) |
| $a/h_0 = 0.3$ (effective length / $h_0 = 10.3$) | | | | |
| 1 | 9.5% (5.4%) | 14.4% (8.7%) | 17.3% 11.5%) | 21.3% (14.4%) |
| 2 | 18.1% (10.4%) | 26.6% (16.8%) | 31.3% (21.3%) | 37.9% (25.9%) |
| $a/h_0 = 0.4$ (effective length / $h_0 = 9.1$) | | | | |
| 1 | 10.1% (5.6%) | 15.2% 9.3%) | 17.7% (11.9%) | 20.3% (14.4%) |
| 2 | 19% (10.7%) | 27.5% (17.5%) | 31.7% (21.9%) | 37.2% (26.6%) |

**Acknowledgements:** The authors are grateful to A.G. Charalampopoulos for providing the code for the calculation of the force on the wall.

# Appendix A. Generation of the solitary wave

As mentioned in the main part of the present paper, to construct an exact, fully nonlinear, initial flow for a water-wave problem, we need to formulate and solve another, hopefully simpler, problem. For solitary waves propagating over a horizontal bottom, the simplification comes from the fact that the problem can be reformulated as a steady (time-independent) one, in terms of a reference frame moving with the constant phase speed $c$ of the waves. In this appendix, we give a brief overview of the highly accurate, iterative solver of Clamond and Dutykh (2013) and Dutykh and Clamond (2014), which has been used for deriving all the initial conditions needed for the calculations of this work.

The solution method goes as follows. The full 2D Euler equations in water of constant depth $d$, assuming that the free-surface elevation is localized near the origin and tends to zero at both infinities ($\pm\infty$), are reformulated as a variational principle with respect to the action functional

$$S[\eta] = \int_{-\infty}^{+\infty} \left[\frac{1}{2}c^2\eta\,C[\eta] - \frac{1}{2}g\eta^2(1+C[\eta])\right]d\alpha, \qquad (20)$$

where $\eta = \eta(\alpha)$ is the free-surface elevation (assumed time-independent in the appropriate reference frame), and $C$ is an appropriate nonlocal (pseudo-differential) operator, taking care of the substrate kinematics. Rendering the action functional $S$ stationary, yields the Euler-Lagrange equation

$$\delta S = c^2 C[\eta] - g\eta - \frac{1}{2}g\,C[\eta^2] - g\eta C[\eta] = 0, \qquad (21)$$

which is known as the Babenko equation for gravity solitary waves (Babenko 1987). The latter is re-written, by separating the linear from the nonlinear part, in the form

$$\mathcal{L}[\eta] = \mathcal{N}[\eta], \qquad (22)$$

where

$$\mathcal{L}[\eta] \equiv c^2\eta - gC^{-1}[\eta] \quad \text{and} \quad \mathcal{N}[\eta] \equiv gC^{-1}[\eta C[\eta]] + \frac{1}{2}g\eta^2. \qquad (23)$$

Eq. (22) is solved by using Petviashvili's iterations (see, e.g., Petviashvili (1976) and other references in the works of Clamond and Dutykh)

$$\eta_{n+1} = S_n^2 \mathcal{L}^{-1}[\mathcal{N}[\eta_n]], \qquad S_n = \frac{\int_{-\infty}^{+\infty}\eta_n \mathcal{L}[\eta_n]d\alpha}{\int_{-\infty}^{+\infty}\eta_n \mathcal{N}[\eta_n]d\alpha}, \qquad (24)$$

where $S_n$ is a *stabilization* factor. The calculations are efficiently performed in the Fourier domain, and communicated to the physical domain by means of the Fast Fourier Transform (FFT). The iteration process uses as initial guess the KdV solution, in the form

$$\eta_0(\alpha) = d(F^2-1)\mathrm{sech}^2(\kappa\alpha/2), \qquad (25)$$

where $\kappa$ is calculated from the equation $F^2 = \kappa d \tan(\kappa d)$, given the Froude number $F \equiv c/\sqrt{gd}$ of the desired solitary wave. The iterations stop when the following criteria

$$\left\|\eta_{n+1}-\eta_n\right\|_\infty < \varepsilon_1, \qquad \left\|\mathcal{L}[\eta_n]-\mathcal{N}[\eta_n]\right\|_\infty < \varepsilon_2 \qquad (26)$$

are met. The tolerance parameters $\varepsilon_1$, $\varepsilon_2$ can be as small as the machine accuracy. Let it be noted that this solver is valid for solitary waves with amplitude-to-depth ratio up to 0.796; see Sec. 3.1 of Clamond and Dutykh (2013).

To use this (fully nonlinear) solitary wave as initial condition for our calculations, we have to appropriately fit it into our computational domain, ensuring the "numerical smoothness" of the initial flow and its validity (i.e. the placement of the solitary wave over an horizontal part of the seabed). Although the solitary wave is theoretically characterized by an infinite wavelength, it is essentially a localized disturbance, which becomes negligible after a finite distance from the wave peak. Accordingly, Clamond and Dutykh present their numerical solution within a finite computational domain, of dimensionless length $l_d \equiv l/d$, centered around the peak of the wave, and sufficiently large so that the free-surface elevation at the ends of this domain is zero up to machine accuracy. Indicatively, the value of $\eta$ at the end points of the computational domain (distance $l_d/2$ from the peak) is of order $O(10^{-16})$, while the value of $\eta$ at distance $l_d/4$ from the peak is of order $O(10^{-9})$. In our simulations, the peak of the solitary wave is placed over a region of uniform depth extending for at least $l_d/4$ from the peak location, in both directions, before encountering any trench or wall. This choice ensures a "numerically smooth" interpolation of the solitary wave solution as a part of the initial flow, used in our simulations.

**Appendix B. Deviation of numerical from experimental results, for the cases considered in Sec. 4.1 and 4.2**

In the following table, the deviations of numerical predictions, $R_{num}$, from the corresponding experimental values, $R_{exp}$, concerning the two peaks of the transformed solitary wave over a shelf and over a step, are summarized in percentages, calculated as

$$\frac{R_{num} - R_{exp}}{R_{exp}} 100\,\%.$$

The considered numerical methods are those appearing in Fig. 2 and Fig. 4 (Sec. 4), and the percentage discrepancies refer to the peak overestimation and time advancement, for each of the two peaks of the transformed solitary wave, at the gauges $g_3$ and $g_4$.

**Table 3** Deviation of various numerical methods from the corresponding experimental results, for the transformation of a solitary wave over a shelf and over a step considered in Sec. 4

| Numerical method | Deviation from Experimental Results | | | |
|---|---|---|---|---|
| | 1st peak (main) | | 2nd peak (secondary) | |
| | overshoot | time advancement | overshoot | time advancement |
| *propagation over a shelf* | | | | |
| **gauge** $g_3$ | | | | |
| HCMT | 14.13% | 1.64% | 52.80% | 2.25% |
| Madsen & Mei, 1969 | 31.65% | - | 82.35% | - |
| Li et al., 2012 | 8.05% | 0.82% | 75.53% | 3.62% |
| **gauge** $g_4$ | | | | |
| HCMT | 36.39% | 1.62% | 49.87% | 1.97% |
| Madsen & Mei, 1969 | 49.61% | - | 95.02% | - |
| Li et al., 2012 | 22.07% | 0.43% | 54.17% | 0.56% |
| *propagation over a step* | | | | |
| **gauge** $g_3$ | | | | |
| HCMT | 18.31% | 3.03% | 3.07% | 2.65% |
| Liu & Cheng, 2001 | 12.13% | 2.23% | 5.21% | 1.34% |
| Seabra-Santos et al., 1987 | 4.15% | 2.60% | 21.17% | -0.16% |
| **gauge** $g_4$ | | | | |
| HCMT | 20.12% | 1.96% | 10.40% | 1.74% |
| Liu & Cheng, 2001 | 12.44% | 1.45% | 0.89% | 0.74% |
| Seabra-Santos et al., 1987 | 4.20% | 1.25% | 20.00% | -1.30% |